# Study of the X-ray Pulsar XTE J1946+274 with *Nustar*

**A. S. Gorban[1,2*], S. V. Molkov[1], S.S.Tsygankov[3,1], and A. A. Lutovinov[1,2]**

*(1) Space Research Institute, Russian Academy of Sciences, Profsoyuznaya ul. 84/32, Moscow, 117997 Russia*

*(2) Higher School of Economics, National Research University, Myasnitskaya ul. 20, Moscow, 101100 Russia*

*(3) Tuorla Observatory, University of Turku, Turku, Finland*

ABSTRACT: We present the results of our spectral and timing analysis of the emission from the transient X-ray pulsar XTE J1946+274 based on the simultaneous NuSTAR and Swift/XRT observations in the broad energy range 0.3–79 keV carried out in June 2018 during a bright outburst. Our spectral analysis has confirmed the presence of a cyclotron absorption line at an energy $\sim 38$ keV in both averaged and phase-resolved spectra of the source. Phase-resolved spectroscopy has also allowed the variation in spectral parameters with neutron star rotation phase, whose period is $\simeq 15.755$ s, to be studied. The energy of the cyclotron line is shown to change significantly (from $\simeq 34$ to $\simeq 39$ keV) on the scale of a pulse, with the line width and optical depth also exhibiting variability. The observed behavior of the cyclotron line parameters can be interpreted in terms of the model of the reflection of emission from a small accretion column (the source's luminosity at the time of its observations was $\sim 3 \times 10^{37}$ erg s$^{-1}$) off the neutron star surface. The equivalent width of the iron line has been found to also change significantly with pulse phase. The time delay between the pulse and equivalent width profiles can be explained by the reflection of neutron star emission from the outer accretion disk regions.

* E-mail: <gorban@iki.rssi.ru>

## 1. INTRODUCTION

The transient X-ray pulsar XTE J1946+274 was first detected during its outburst in September 1998 with the All-Sky Monitor (ASM) onboard the Rossi X-ray Timing Explorer (*RXTE*, orbital observatory (Smith and Takeshima 1998a, 1998b). Subsequently, coherent pulsations with a period of 15.83 s were recorded from the source based on data from the BATSE instrument onboard the *Compton-GRO* observatory (Wilson et al. 1998). Three years after its discovery the pulsar went into quiescence and exhibited no outburst activity in X-rays until June 2010, when the Burst Alert Telescope (BAT) onboard the Gehrels*Swift* observatory and the Gamma-ray Burst Monitor (GBM) onboard the *Fermi* observatory recorded a new outburst (Krimm et al. 2010; Finger 2010). During this outburst the source was also observed with the *INTEGRAL*, *RXTE* and *Suzaku* orbital observatories (Caballero et al. 2010; Muller et al. 2012; Maitra and Paul 2013; Marcu-Cheatham et al. 2015). As in the case of the 1998 outburst, several fainter events, type I outbursts, associated with the binary periastron passage by the neutron star were observed after the main type II outburst (bright outbursts in Be systems independent of the binary orbital phase with a peak luminosity reaching the Eddington limit for a neutron star) (Muller et al. 2012), whereupon the emission from the source again was not recorded up until 2018, when *Fermi*/GBM recorded another outburst from XTE J1946+274 (Jenke et al. 2018).

The optical companion in the system was established owing to a good localization in X-rays based on *BeppoSAX*, data, which allowed its spectroscopic study to be carried out in the optical range (Verrecchia et al. 2002). Strong $H_\alpha$ and $H_\beta$ emission lines were detected in the spectra of the optical companion of the pulsar. Spectroscopic and photometric data made it possible not only to determine the spectral type of the companion star (B0-1V-IVe), but also to estimate the distance to the binary system, $d \sim 8 - 10$ kpc (Verrecchia et al. 2002). Thus, the set of X-ray and optical data allows the source to be assigned to the class of X-ray pulsars in binary systems with Be stars (BeXRBs). Subsequently, while analyzing the correlation between the neutron star spin-up rate and observed flux, Wilson et al. (2003) estimated the distance to the pulsar to be $d = 9.5 \pm 2.9$ kpc (this value will be used below). A large distance to the system $\sim 10$ kpc was also confirmed by data from the *Gaia* telescope (Arnason et al. 2021).

As has been said above, the first outburst of XTE J1946+274 in 1998 lasted about three months. After the main outburst the source had remained active for approximately three more years, exhibiting a series of fainter bursts separated by $\sim$ 80 days (Campana et al. 1999). Subsequently, this periodicity was shown to be associated with the neutron star motion in the orbit of the binary system with a period of 169.2 days (see Wilson et al. 2003). The orbital parameters of the binary system were first determined in the same paper and were subsequently refined by Marcu-Cheatham et al. (2015).

In quiescence the source was observed in March 2013, several years after the 2010 outburst, with the *Chandra* observatory and was detected in a state with a luminosity $10^{34}$ erg $s^{-1}$ (Ozbey Arabaci et al. 2015; Tsygankov et al. 2017a). Interestingly, despite such a low luminosity, pulsed emission was recorded from XTE J1946+274 and its hard energy spectrum was measured (Tsygankov et al. 2017a), suggesting ongoing accretion even at such low luminosities, which is most likely due to the presence of a weakly ionized (cold) accretion disk around the neutron star (Tsygankov et al. 2017b).

In its bright state the energy spectrum of XTE J1946+274 has a shape typical of X-ray pulsars and can be described by a power law with a high-energy exponential cutoff. In addition, using *RXTE*, data, Heindl et al. (2001) managed to detect a cyclotron absorption line at an energy $\sim$35 keV, which allowed the magnetic field strength of the neutron star in the system to be measured. Subsequently, the presence of a cyclotron line at an energy $\sim$38 keV in the source's spectrum during its 1998 outburst was independently confirmed by *BeppoSAX* data (Doroshenko et al. 2017). However, while analyzing the *RXTE*, data obtained during the 2010 outburst, Muller et al. (2012) ruled out the presence of a cyclotron line at 35 keV. Instead, the authors found evidence for the presence of an absorption feature at an energy $\sim$25 keV. This gave the authors grounds to assume that the cyclotron energy could vary from outburst to outburst. At the same time, based on the *Suzaku*data obtained during the same outburst, Maitra and Paul (2013) and Marcu-Cheatham et al. (2015) confirmed the presence of an absorption line at energies 35–38 keV and found no signatures of such a feature near 25 keV.

Thus, the question about the presence and accurate parameters of the cyclotron feature in the spectrum of the pulsar XTE J1946+274 still remains open and is discussed in this paper. To answer this question, we use the *Nustar* data obtained when observing the pulsar in a bright state, during the 2018 outburst.

## 2. OBSERVATIONS AND DATA ANALYSIS

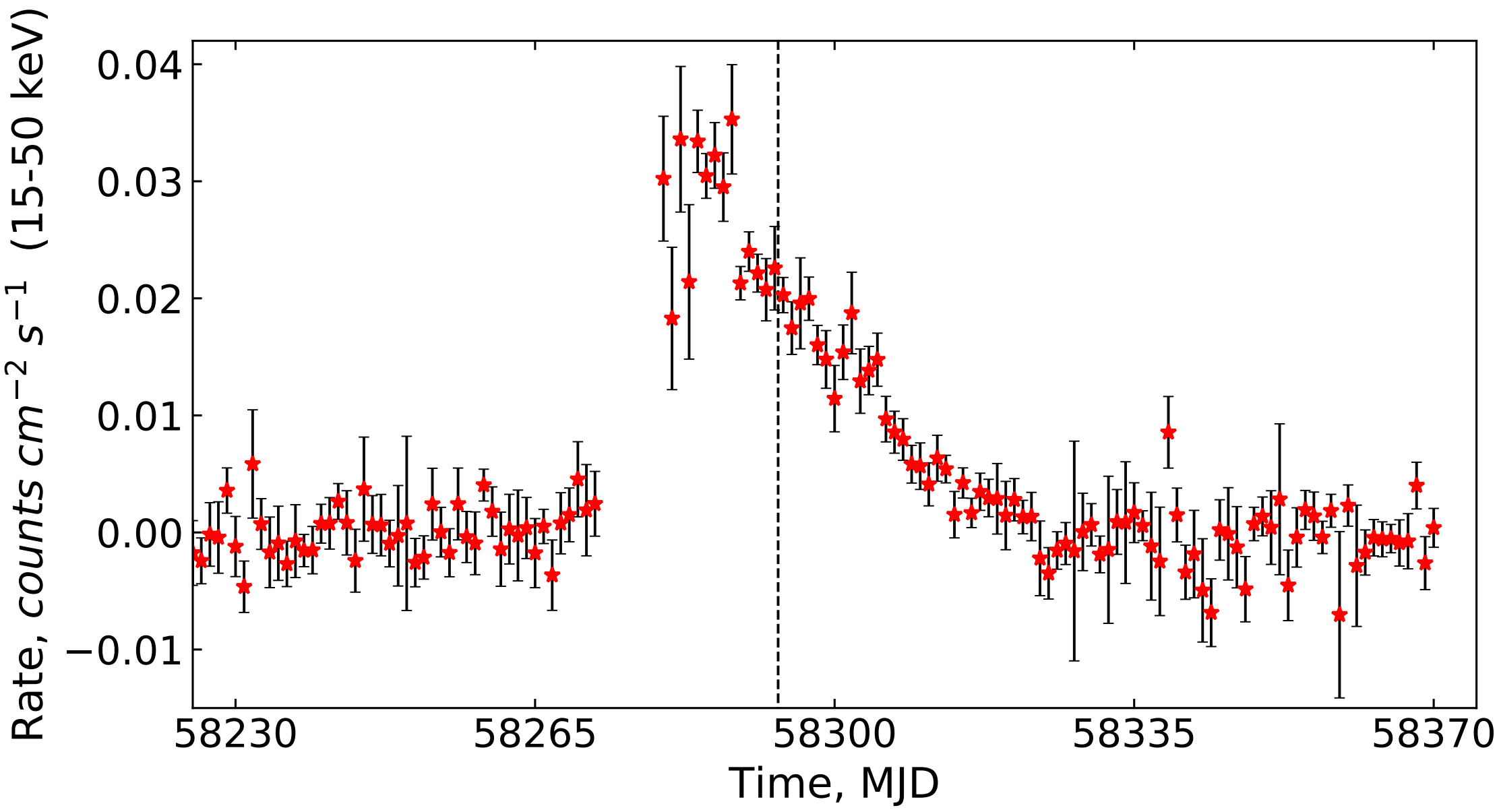

Figure 1: Light curve of the pulsar XTE J1946+274 in the 15–50 keV energy band from *Swift*/BAT data. The dashed line marks the time of *Nustar* observation.

In 2018, after seven years of its stay in quiescence, the sourceXTE J1946+274 went into another activity phase. The onset of a type II outburstwas recorded with the *Fermi*/GBM instrument (Jenke et al. 2018). The outburst lasted more than two months. To show its morphology, Fig. 1 presents the light curve in the 15–50 keV energy band from *Swift*/BAT data (Krimm et al. 2013). The source was observed with several observatories and instruments in the outburst time (NICER, NuSTAR, Swift). However, to study the spectral and timing (*NICER*, *Nustar* *Swift*) characteristics in a wide energy range, we used only the *Nustar* data and the *Swift*/XRT data for a spectral analysis in soft X-rays (0.3–10 keV).

The *Nustar* (Nuclear Spectroscopic Telescope ARray) observatory consists of two identical X-ray telescopes (FPMA and FPMB) operating in the energy range 3–79 keV with an energy resolution ∼400 eV at 10 keV (Harrison et al. 2013). The observations of XTE J1946+274 were carried out on June 24, 2018, (ObsID 90401328002) with an exposure time ∼50 ks (the time of observation is marked in Fig. 1 by the dashed line).

The *Nustar* data were analyzed with the HEASOFT v6.27.2 software package and using the *Nustar* Data Analysis (NuSTARDAS v0.4.7) software and CALDB v20180925 calibrations. After the data reduction with NUPIPELINE, we used NUPRODUCTS to extract the energy spectra of the source and its light curves. The data for the source were extracted from a circular region 100" in radius located symmetrically relative to the source image center on the detectors. Since the source is very bright, the region for extracting the spectrum and the light curve of the background, whose size is 150", was chosen on the adjacent chip. For our spectral analysis all of the extracted energy spectra were binned by 25 counts per bin.

The observations of XTE J1946+274 were carried out with *Swift*/XRT simultaneously with the *Nustar* observations on June 24, 2018 (ObsID 00088783002) with an exposure time of 1.2 ks. The spectra of the source were obtained with the online tools (Evans et al. 2009) provided by the UK Swift Science Data Centre at the University of Leicester.[1] We used the WT mode data, because the PC mode data were subject to pile-up, which do not allow the results to be adequately assessed. The energy spectra were also binned by 25 counts per bin.

## 3. RESULTS

### *3.1. Timing Analysis of the Emission from XTE J1946+274 during the 2018 Outburst*

First we performed a detailed timing analysis of the . To take into account the effects associated with the motion of the spacecraft around the Earth and of the Earth around the Sun, first of all, all of the photon arrival times were reduced to the Solar system barycenter. Barycentering was done with standard tools from the HEASOFT software package. No correction of the photon arrival time for the orbital motion in the binary system was made, because the orbital period is much longer than the time of observations and, therefore, this correction could be neglected for the purposes of this paper. Thus, we determined the pulsation period, $15.75519 \pm 0.00001$ s. The standard epoch-folding technique was used in the search for a periodic signal. The error in the period was estimated using Monte Carlo simulations of the light curve (Boldin et al. 2013). The next step was the light curve extraction from the data of each module in the 3–6, 6–8, 8–10, 10–15, 15–20, 20–30, 30–40, and 40–78 keV energy bands. The light curves were constructed with a time resolution of 0.1 s. Then, we subtracted the background from the source's light curves and combined the light curves of the two modules with the lcmath tool (FTOOLS V6.27). The pulse profile was reconstructed by folding the light curve with the pulsation period determined above. Figure 2 presents the derived phase light curves of the source. It can be clearly seen that the pulse profile changes significantly with energy. Approximately up to 20 keV the pulse profiles are similar in overall structure and are characterized by two phase regions—they have a double peak with a fairly narrow minimum between them at phase 0.6. At the same time, the main peak at low energies (up to $\sim$ 10 keV) exhibits a complex structure. At higher energies the main peak virtually disappears and one peak remains at phases from 0.8 to 1.0. To study in more detail the evolution of the emission as a function of energy, we constructed a two-dimensional distribution of pulse profile intensities (Fig. 3). As in Fig. 2, the change of the profile with energy from double-peaked to single-peaked is traceable here, with this transition being observed near the cyclotron line energy. Such transitions from a double-peaked structure to a single-peaked one have been detected previously and were described in Wilson et al. (2003) during the two outbursts observed with *RXTE*/PCA in 1998 and 2001 and in Doroshenko et al. (2017) based on the *BeppoSAX* data obtained in 1998. Interestingly, such a behavior has also been observed previously in other BeXRB systems (Tsygankov et al. 2006, 2007, 2015; Iyer et al. 2015) and is explained as a consequence of the effects of an angular redistribution of X-ray emission due to cyclotron resonance scattering in a strong magnetic field in combination with relativistic effects and the geometry of the emitting region (see, e.g., Ferrigno et al. 2011; Schonherr et al. 2014).

[1] http://www.swift.ac.uk/user_objects/

In Fig. 4 the pulsed fraction is plotted against energy. The puled fraction was defined as the ratio (Fmax - Fmin)/(Fmax + Fmin), where Fmax and Fmin are the maximum and minimum fluxes in the pulse profile (using 10 phase bins in all energy channels). It can be seen from this plot that the pulsed fraction increases with energy, which is typical of most X-ray pulsars, especially bright ones (Lutovinov and Tsygankov 2009). Doroshenko et al. (2017) showed that there is a feature near an energy $\sim$ 40 keV associated with the local increase in pulsed fraction near the cyclotron absorption line detected previously for several other X-ray pulsars (Tsygankov et al. 2007; Lutovinov and Tsygankov 2009; Lutovinov et al. 2016). On the whole, the *Nustar* results (Fig. 4) agree well with the *BeppoSAX* ones. However, no clear increase in pulsed fraction is recorded near the cyclotron line.

### *3.2. Spectral Analysis of the Source XTE J1946+274 during the 2018 Outburst*

The averaged spectrum of XTE J1946+274, obtained from the *Nustar* and *Swift*/XRT data is shown in Fig. 5a. Two models commonly used to describe the spectra of X-ray pulsars were applied to describe the continuum in the XSPEC v12.11.0 package (Arnaud et al. 1999): a power law with a high-energy exponential cutoff (POW-

Table 1: Parameters of the spectrum for XTE J1946+274 with the continuum described by the CompTT+BB and HIGHECUT+BB

| Model parameters | CompTT+BB | HIGHECUT+BB |
|---|---|---|
| $N_H$, $10^{22}$ cm$^{-2}$ | $0.14^{+0.25}_{-0.13}$ | $0.86^{+0.29}_{-0.31}$ |
| $\Gamma$ | – | $0.92^{+0.03}_{-0.03}$ |
| $kT_{bb}$, keV | $1.59^{+0.02}_{-0.02}$ | $2.14^{+0.12}_{-0.09}$ |
| $E_{cut}$, keV | – | $18.02^{+0.17}_{-0.17}$ |
| $E_{fold}$, keV | – | $8.92^{+0.18}_{-0.17}$ |
| $E_{Fe}$, keV | $6.44^{+0.03}_{-0.03}$ | $6.44^{+0.03}_{-0.03}$ |
| $\sigma_{Fe}$, keV | $0.42^{+0.04}_{-0.03}$ | $0.40^{+0.04}_{-0.03}$ |
| $W_{Fe}$, keV | $0.116^{+0.006}_{-0.002}$ | $0.107^{+0.005}_{-0.004}$ |
| $E_{Cycl}$, keV | $37.49^{+0.70}_{-0.64}$ | $37.81^{+0.75}_{-0.73}$ |
| $\sigma_{Cycl}$, keV | $8.59^{+0.69}_{-0.61}$ | $4.53^{+0.59}_{-0.55}$ |
| $\tau_{Cycl}$ | $0.567^{+0.011}_{-0.011}$ | $0.248^{+0.003}_{-0.003}$ |
| $T_{0Comptt}$, keV | $0.46^{+0.13}_{-0.46}$ | – |
| $kT_{Comptt}$, keV | $7.07^{+0.07}_{-0.06}$ | – |
| $\tau_{Comptt}$ | $6.39^{+0.07}_{-0.06}$ | – |
| $Flux$ (3 - 79 keV), $10^{-9}$ $erg\ cm^{-2}\ s^{-1}$ | $2.603^{+0.005}_{-0.065}$ | $2.606^{+0.001}_{-0.011}$ |
| $C_{NuSTAR}$ | $1.005^{+0.002}_{-0.002}$ | $1.005^{+0.002}_{-0.002}$ |
| $C_{XRT}$ | $0.719^{+0.081}_{-0.079}$ | $0.710^{+0.076}_{-0.075}$ |
| $\chi^2$ (d.o.f.) | 2083.95 (1901) | 2065.27 (1900) |

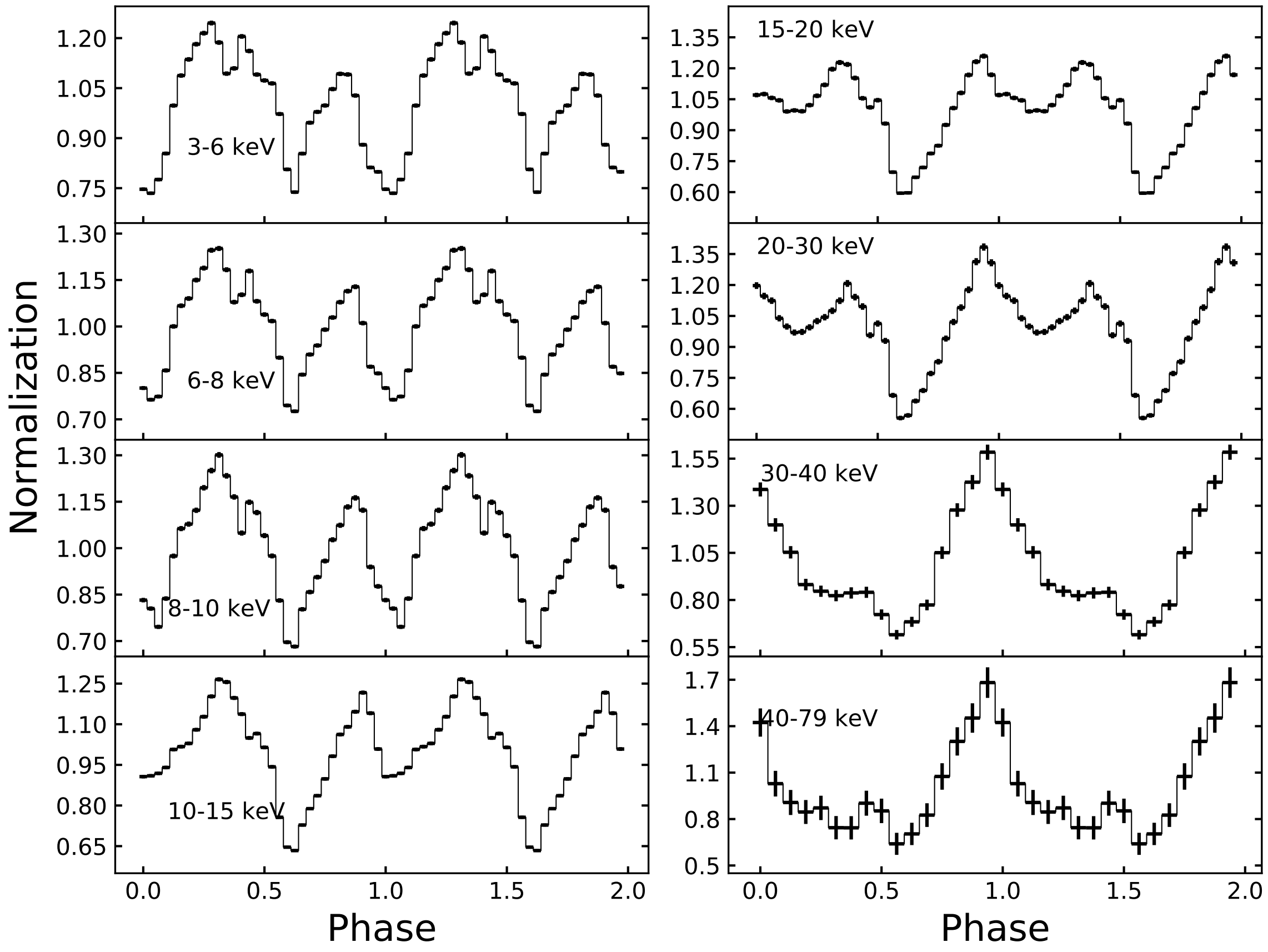

Figure 2: Pulse profiles for XTE J1946+274 in eight energy bands from the NuSTAR data in June 2018. The count rate was normalized to the mean value in a given band.

ERLAW*HIGHCUT) and a Comptonization model (COMPTT, Titarchuk 1994). The TBABS component was added to take into account the interstellar absorption. In addition, the iron emission line at ~6.4 keV was added to the combined best-fit model. The iron line equivalent width $W_{Fe}$ for the averaged spectrum is $\simeq$ 0.1 keV. The spectra for the two *Nustar* modules and *Swift*/XRT were fitted simultaneously. To take into account the various calibrations of these modules and the nonsimultaneity of the *Nustar* and *Swift* observations, we introduced normalization coefficients (see Table 1), while the remaining model parameters were fixed between themselves. Our analysis revealed that the COMPTT and POWERLAW*HIGHCUT models fit the data approximately identically, with slight differences.

In particular, the TBABS*(BB+COMPTT+GAUS) model shows an unsatisfactory quality of the fit with $\chi^2 = 2716.94$ with 1904 degrees of freedom and noticeable deviations, implying a deficit of photons, near ∼38 keV (Fig. 5b). To describe this feature, an absorption line with a Gaussian optical depth profile GABS, which can be interpreted as the cyclotron resonance scattering line, was added to the model. This led to a significant improvement in the quality of the model fit, $\chi^2 = 2083.95$ (1901) and an adequate description of the source's spectrum (Fig. 5c). The width of the cyclotron line and its optical depth are $\sigma_{Cycl} \simeq 8.6$ keV и $\tau_{Cycl} \simeq 0.57$, respectively. The observed X-ray flux from XTE J1946+274 in the energy range 3–79 keV is $F_x = (2.60 \pm 0.05) \times 10^{-9}$ erg cm$^{-2}$ s$^{-1}$, corresponding to a luminosity $L_x \simeq 2.8{\times}10^{37}$ erg s$^{-1}$ for a distance of 9.5 kpc. Table 1 also gives the derived continuum parameters for the averaged spectrum described by the COMPTT model.

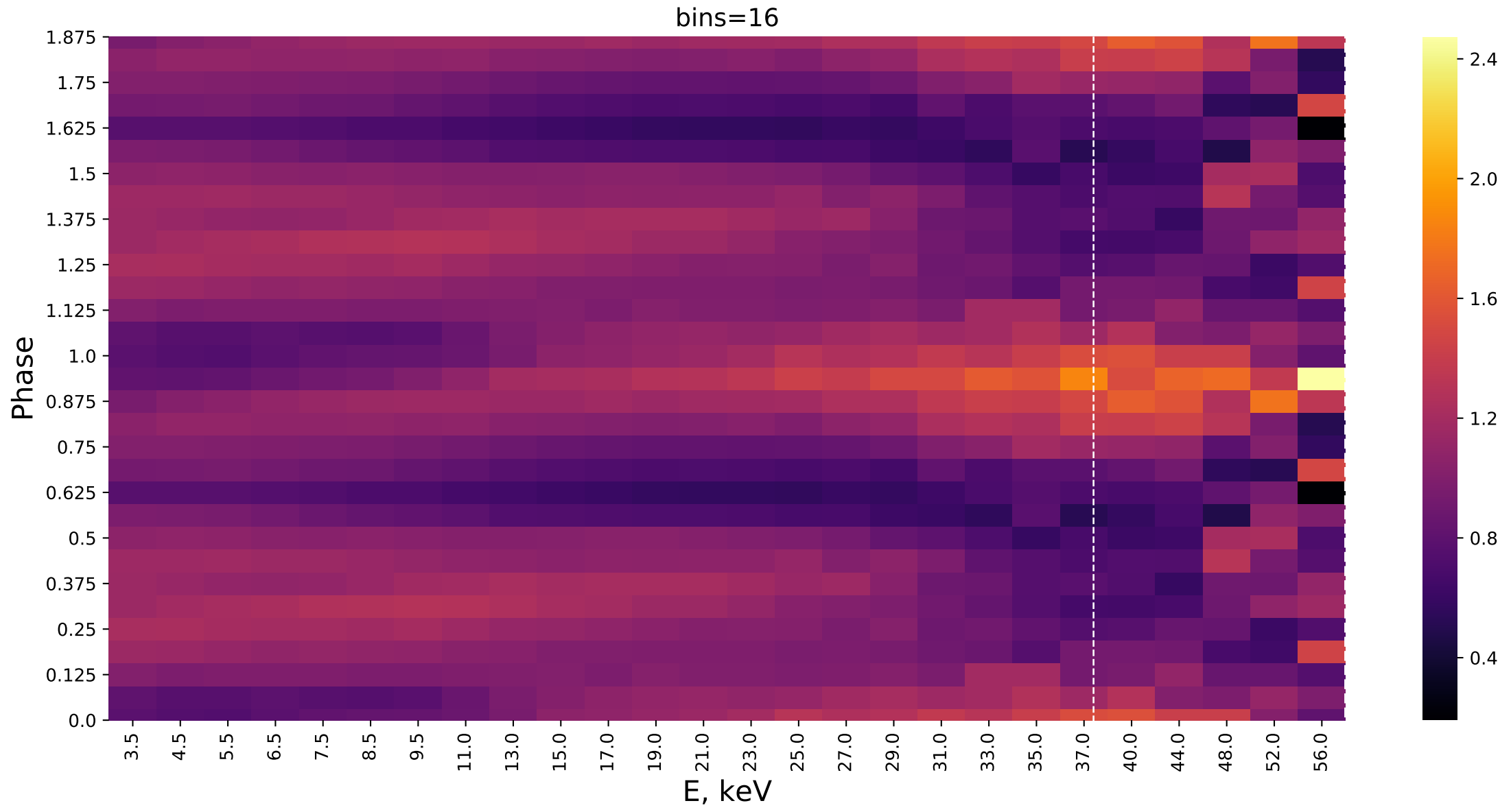

Figure 3: Two-dimensional distribution of the normalized pulse profile intensity as a function of energy. Lines of the same color indicate equal values of the normalized intensity presented on the right panel. The dashed line indicates the position of the cyclotron absorption line centroid in the pulsar spectrum (see the Subsection "Spectral Analysis").

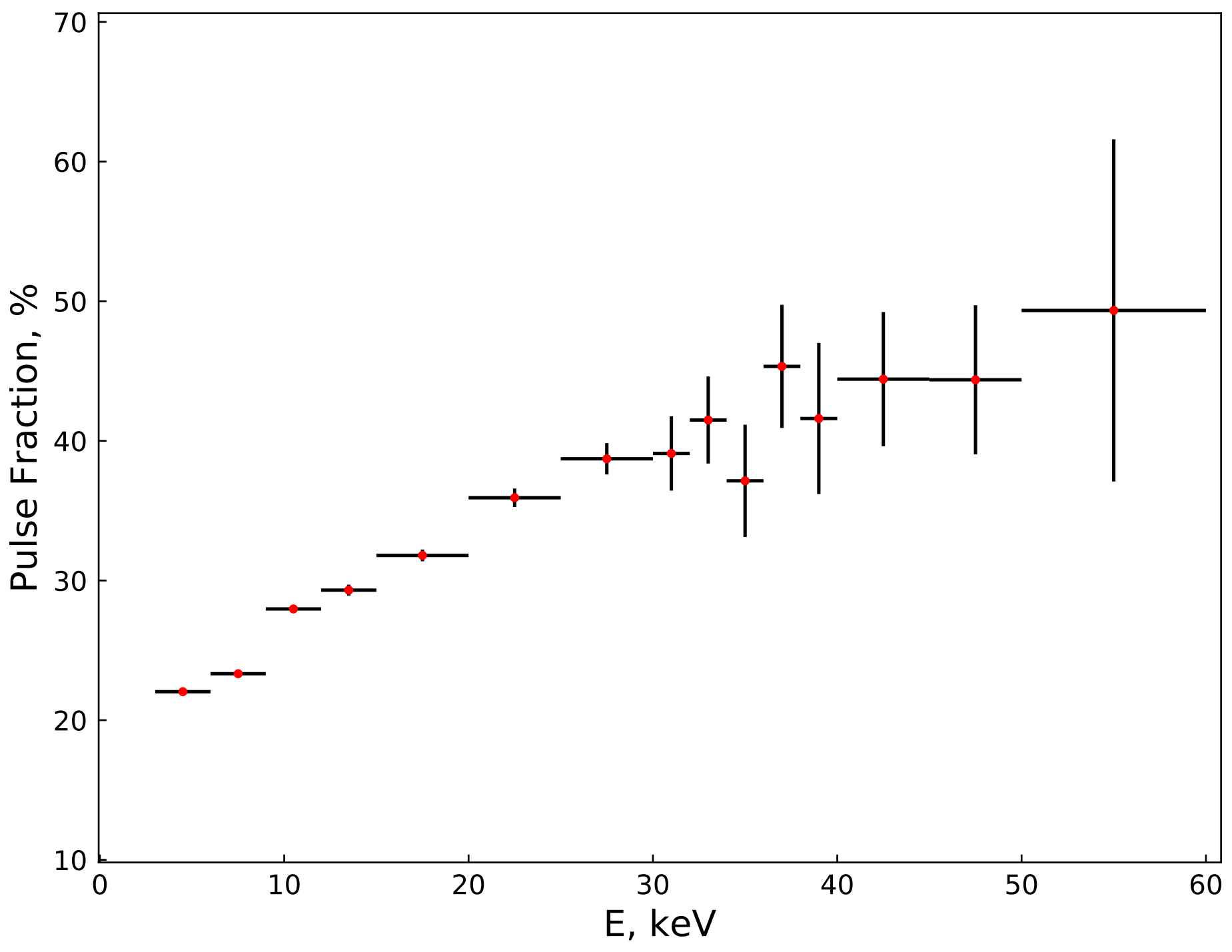

Figure 4: Pulsed fraction for XTE J1946+274 versus energy from the *Nustar* data.

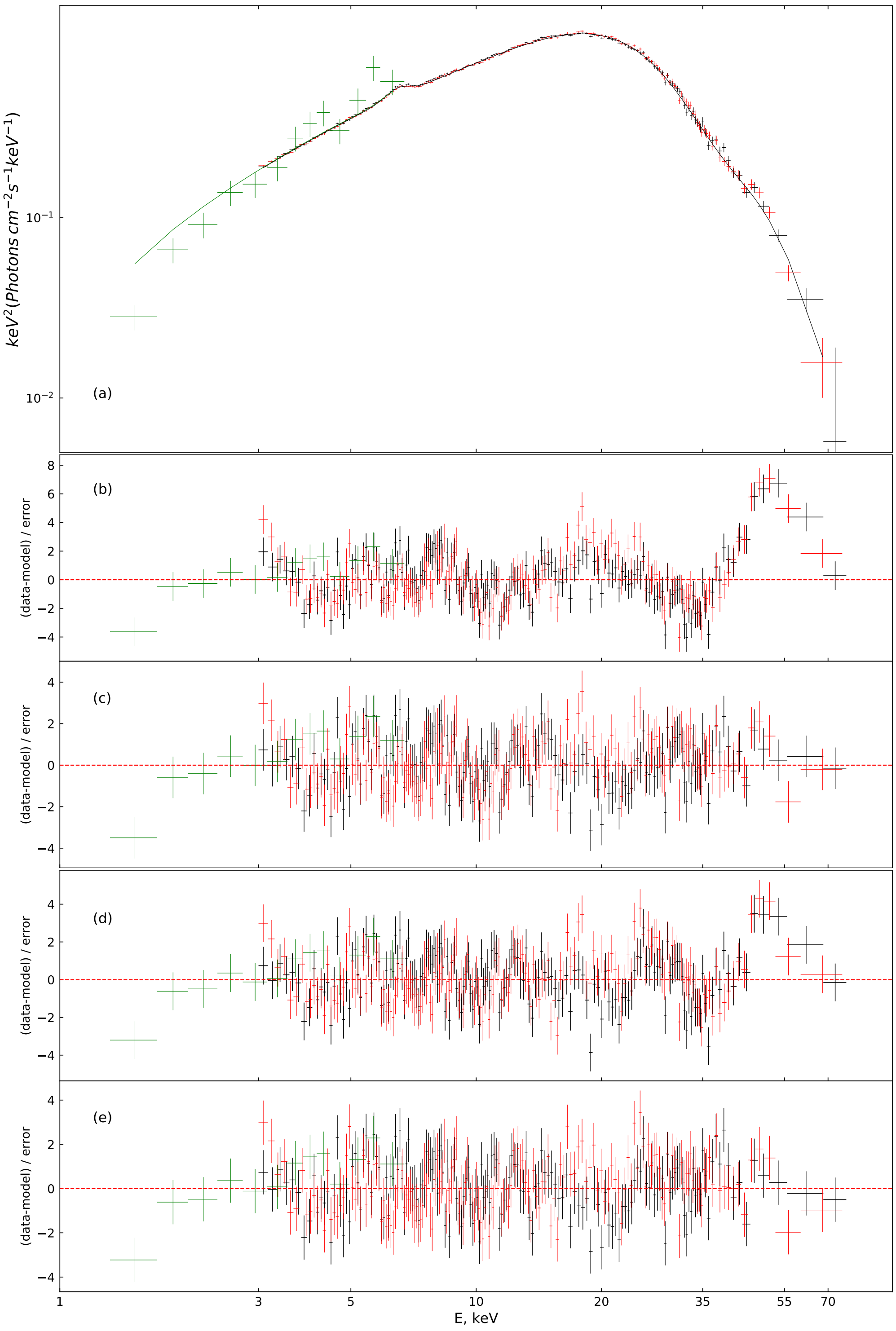

Figure 5: (a) The energy spectrum of XTE J1946+274 measured from the *Nustar* (red and blue dots) and *Swift*/XRT (green dots) data; the solid line indicates the best-fit model for the spectrum. Panels (b) and (c) show the deviation of the observational data from the HIGHECUT+BB model without and with the inclusion of the cyclotron line in the model, respectively; panels (d) and (e) show the deviation of the observational data from the CompTT+BB model without including the cyclotron line and with the inclusion of all the additional model components, respectively.

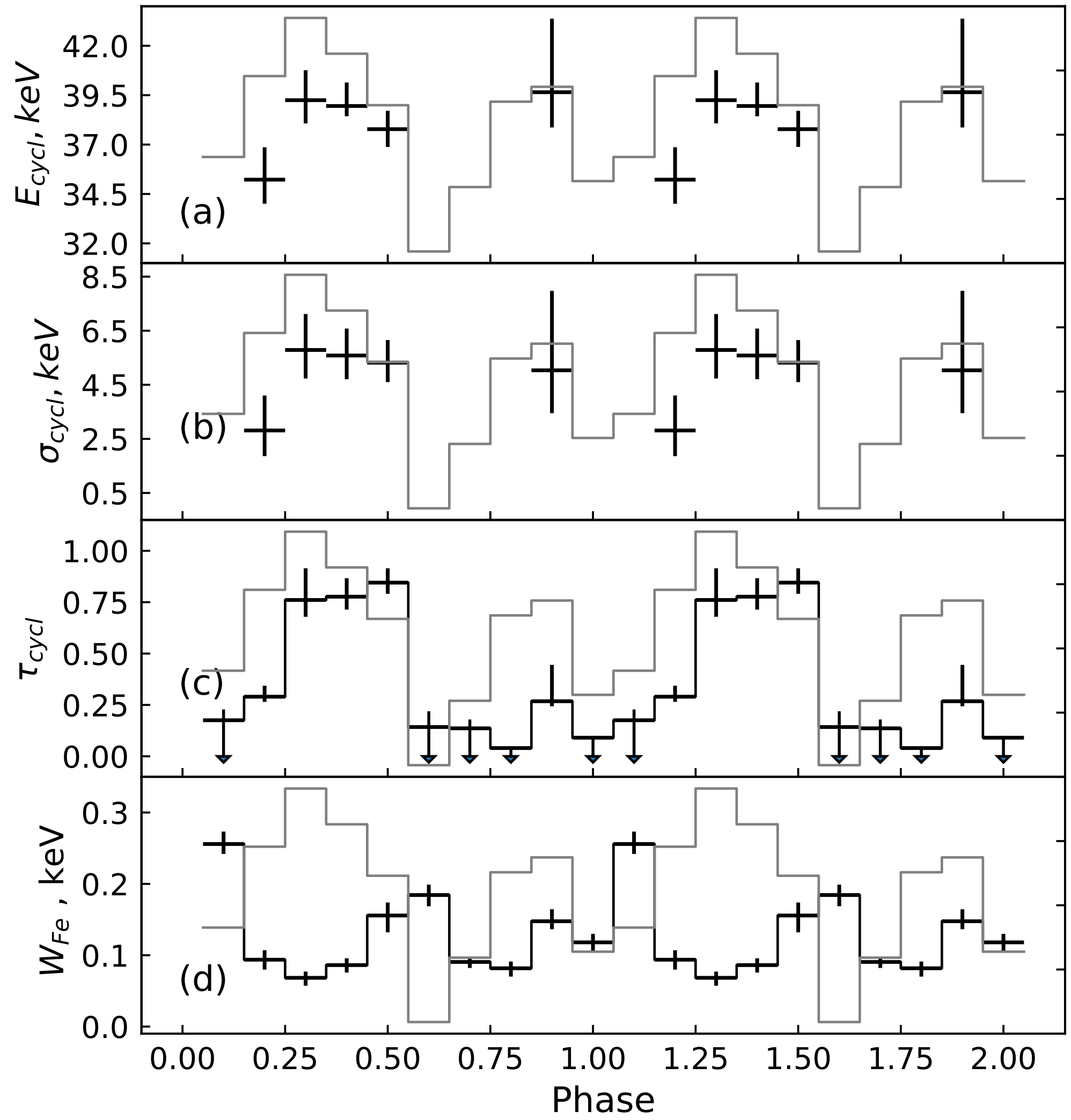

Figure 6: Spectral parameters of XTE J1946+274 versus pulse phase: (a) the cyclotron energy, (b) the cyclotron line width, (c) the cyclotron line optical depth, and (d) the equivalent width of the iron emission line. The gray line indicates the pulse profile in the complete energy range of the *Nustar* observatory.

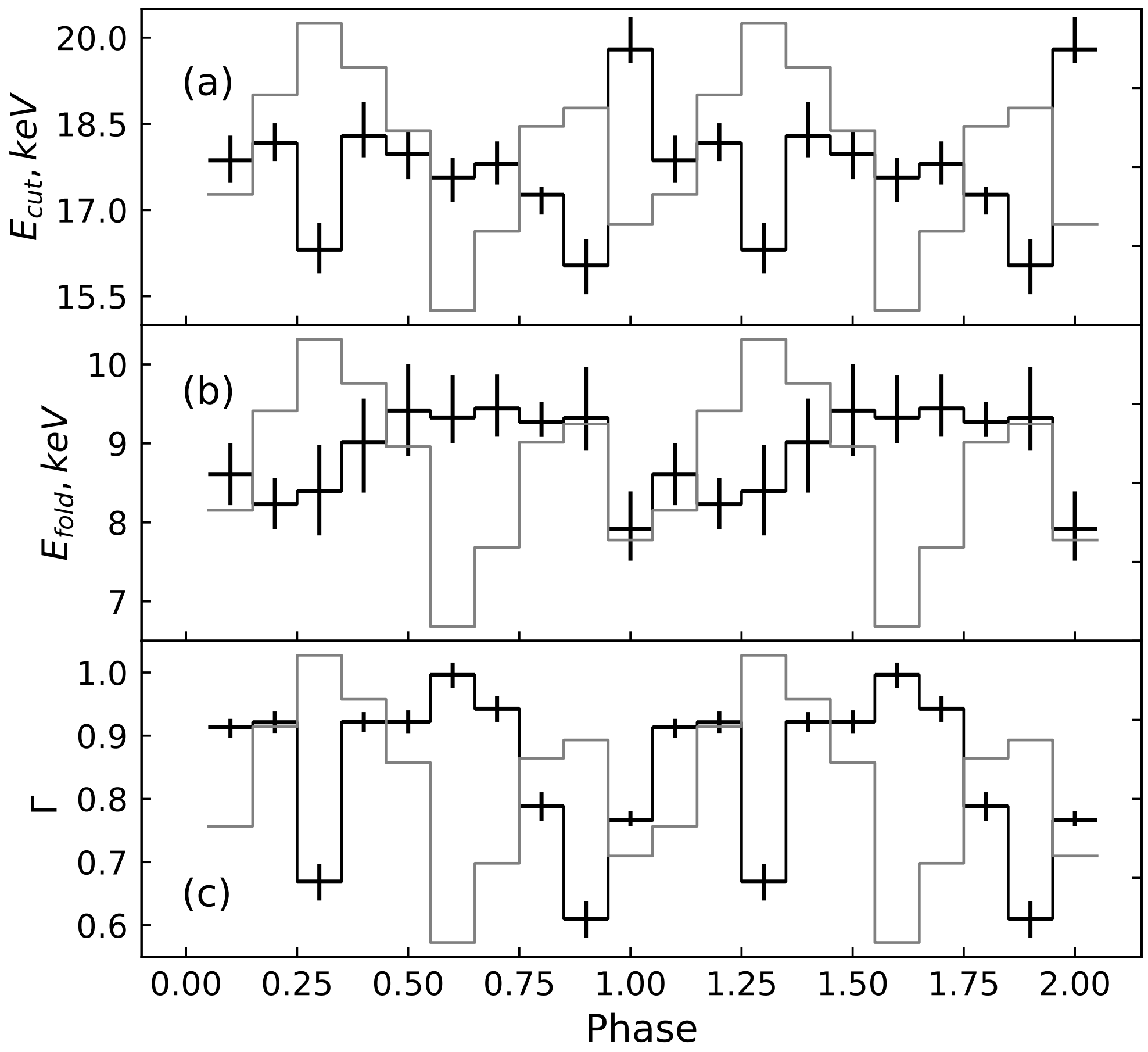

Figure 7: Spectral continuum parameters for XTE J1946+274 versus pulse phase: (a) the cutoff energy, (b) the exponential decay energy, and (c) the photon index.

To compare our results with previous studies (Marcu-Cheatham et al. 2015; Doroshenko et al. 2017), the spectrum was also fitted by the TBABS*(GAUS+BB+POWERLAW*HIGHCUT) model and showed a quality of the fit equal to 2222.91 per 1903 degrees of freedom.[2]. Just as in the TBABS*(BB+COMPTT+GAUS) model, a feature that was described by adding the GABS component to the main model is observed near an energy ∼38 keV. The quality of the fit improved significantly to 2065.27 (1900) (Figs. 5d and 5e). The width of the cyclotron line is $\sigma_{Cycl} \simeq 4.53$ keV, while its optical depth is $\tau_{Cycl} \simeq 0.25$. The parameters of the POWERLAW*HIGHCUT+BB model for the averaged spectrum are also given in Table 1.

Thus, irrespective of the model describing the continuum, a feature near 38 keV, which is most likely associated with cyclotron absorption, is recorded in the source's spectrum at a statistically significant level. The energy of the line is virtually independent of the choice of a continuum model, while its width and optical depth are considerably smaller for the powerlaw model with an exponential cutoff.

To study the evolution of the spectral parameters as a function of the pulse phase (or, in other words, the viewing angle of the neutron star emitting regions), we performed phase-resolved spectroscopy. Taking into account the observed morphology of the pulse profile, we divided the data into ten phase bins. To fit the spectra, we used the POWERLAW*HIGHCUT+BB model, just as for the averaged spectrum. Since the blackbody temperature is poorly determined in phase-resolved spectra, it was fixed at the value for the averaged spectrum. Our analysis revealed that the iron emission line is present at all phases, while the cyclotron absorption line is recorded at a statistically significant level only in five phase bins with emission peaks. At the same time, it turned out that the energy of the cyclotron line changes significantly, from 34 to 39 keV, on the scale of a pulse (Fig. 6a), reaching its maximum values near the pulse peaks. This phase dependence of the energy is probably related to the changes in the viewing angle of the regions where the cyclotron line is formed (see, e.g., Lutovinov et al. 2015). Moreover, the width and depth of the line also change during the emission pulse in approximately the same way as its energy. The dependences of the continuum parameters on pulse phase are shown in Fig. 7.

[2]It is important to note that in this model an artificial deficit of photons can arise at the energy $E_{cut}$, which is compensated for by the inclusion of an additional absorption line with an energy equal to $E_{cut}$ and a width of $0.1E_{cut}$.

Apart from the changes in cyclotron line parameters, significant changes in the iron line equivalent width are also observed on the scale of a pulse. In this case, the maxima of the equivalent width are shifted with respect to the maxima of the pulse profile (Fig. 6d). A similar behavior has also been observed previously for other pulsars (see, e.g., Tsygankov and Lutovinov 2009; Shtykovsky et al. 2017). This phase shift for XTE J1946+274 can be estimated as $\triangle\phi \backsim 0.8$. To determine the phase shift, we performed a cross-correlation, during which we determined the most probable phase deviation. Thus, the above phase shift corresponds to a time delay $\triangle\ t \backsim 12.6$ s at a pulsar spin period $P_{spin} \simeq 15.755$ s. The distance that the photons will travel in 12.6 s is $\backsim 3.78\times10^{11}$ cm. At the same time, the outer radius of the accretion disk has an approximate value between the circularization radius $\backsim 1.1\times10^{12}$ cm (the radius at which the angular momentum of the material in a circular orbit is equal to the angular momentum of the material transferred from the Lagrange point L1; Hasayaki and Okazaki 2004) and the Roche lobe radius $\backsim 5.6\times10^{12}$ cm, with the inner boundary of the accretion disk having a radius $\backsim 6\times10^{8}$ cm. The distance corresponding to the delay between the maxima of the pulse profile and the equivalent width may roughly correspond to the accretion disk radius if the reflection occurs at its outer edge, which may have warped regions (Shtykovsky et al. 2017). The above estimates show that, in this case, the accretion disk either has a small size compared to the Roche lobe (Hasayaki and Okazaki 2004) or this delay may be greater than the above time by a whole number of neutron star spin periods.

## 4. CONCLUSIONS

In this paper we analyzed the observational data for the X-ray pulsar XTE J1946+274, obtained with the *Nustar* observatory in June 2018. We showed that the broadband spectrum of the source could be best described either by the Comptonization model COMPTT, or by a power law with a high-energy exponential cutoff, including the absorption at low energies and the fluorescent iron line at 6.4 keV. In addition, a cyclotron absorption line at an energy $\sim$ 38 keVwas detected in the spectrum(Table 1), confirming the results of previous observations (Wilson et al. 2003; Doroshenko et al. 2017). The measured line energy allows the magnetic field strength on the neutron star surface to be determined, B $\backsim$3.2 $\times 10^{12}$ G.

The observed pulse profiles change noticeably with increasing energy. Two peaks separated approximately by half the phase are observed at energies from 3 to 20 keV. The most natural explanation of this fact is that these two peaks are associated with the emission from the two neutron star poles. As the energy increases, these peaks are transformed into one peak that is observed up to $\backsim$ 79 keV. As has been pointed out above, such a behavior is probably associated with the effect of an angular redistribution of X-ray emission due to cyclotron resonance scattering in a strong magnetic field in combination with relativistic effects and the geometry of the emitting region (Ferrigno et al. 2011; Schonherr et al. 2014). Interestingly, the transition from the double-peaked structure to the single-peaked one occurs near the cyclotron line energy. This observational fact agrees well with similar results found previously for several other X-ray pulsars. To study the behavior of the source and the parameters of its emission on the scale of a single pulse, we performed phase-resolved spectroscopy. The latter showed that the energy and other parameters of the cyclotron feature change significantly with pulse phase. A change in the observed parameters of the cyclotron line on the scale of one neutron star rotation is quite typical of X-ray pulsars, with the relative change in the line energy $\sim$ 30% for XTE J1946+274 being in good agreement with the values measured for several other pulsars (Staubert et al. 2019). The source's luminosity, on which, in turn, the physical picture of the formation of observed emission depends, is of great importance for explaining the observed modulation of the line parameters.

As has been shown above, the pulsar's luminosity during its *Nustar* observation is $L_x$ $\sim$ 3$\times 10^{37}$ erg s$^{-1}$, which is close to the critical luminosity corresponding to the appearance of a radiation-dominated accretion column from both theoretical (Basko and Sunyaev 1976; Becker et al. 2012; Mushtukov et al. 2015) and observational (the pulsar V0332+53, Doroshenko et al. 2017) points of view. Thus, given the similarity between the observed parameters of XTE J1946+274 and V0332+53 (Lutovinov et al. 2015), the presence of an accretion column in both cases can be assumed. This allows the observational changes in cyclotron line parameters to be described in terms of the Poutanen et al. (2013) model, in which the cyclotron absorption line is formed in the emission reflected from the neutron star surface. In this case, interestingly, despite the fact that the source was observed at various luminosities (from $5 \times 10^{36}$ to $5 \times 10^{37}$ erg s$^{-1}$), no changes in the line energy were detected (Marcu-Cheatham et al. 2015; Doroshenko et al. 2017).

Our phase-resolved spectroscopy also showed that the maxima of the iron line equivalent widths do not coincide with the maxima of the pulse profile. This allowed us to determine the time delay ($\backsim$ 12.6 s) between the emission and equivalent width peaks, corresponding to a distance $\backsim 3.8 \times 10^{11}$ cm. This value exceeds the inner size of the accretion disk by more than two orders of magnitude, but, at the same time, is much smaller than the distance to the companion star in the system. Given that the fluorescent iron line results from the reflection of the hard X-ray radiation emitted near the neutron star by a fairly cold material, it can be assumed that its formation region is at the outer boundary of the accretion disk. At the same time, this region must be compact enough to provide the observed variability of the iron line equivalent width with pulse phase.

## 5. ACKNOWLEDGMENTS

This study was carried out using the data obtained with NuSTAR, a Caltech project, funded by NASA and operated by NASA/JPL and the data provided by the UK Swift Science Data Centre (XRT data analysis). In this study we also used the software provided by the High-Energy Astrophysics Science Archive Research Center (HEASARC), which is a service of the NASA/GSFC Astrophysics Science Division.

## 6. FUNDING

This work was supported by RSF grant no. 19-12-00423.